\documentclass[english]{article}
\usepackage[T1]{fontenc}
\usepackage[latin9]{inputenc}
\usepackage{textcomp}
\usepackage{amsmath}
\usepackage{amssymb}
\usepackage{setspace}
\makeatletter

\providecommand{\tabularnewline}{\\}

\newcommand{\lyxaddress}[1]{
	\par {\raggedright #1
	\vspace{1.4em}
	\noindent\par}
}

\makeatother

\usepackage{babel}
\begin{document}
\title{Binomial expansion for a linearized map with memory. A final comment
on: \textquotedblleft Criticality and the fractal structure of -5/3
turbulent cascades\textquotedblright{}}
\author{Juan S. Medina-Alvarez\textsuperscript{a,b,{*}}}
\maketitle

\lyxaddress{\textsuperscript{a}Deeptikus Ltd., 10A Tasman Ave., Mount Albert,
Auckland 1025, New Zealand\\\textsuperscript{b}Avda. Rafael Cabrera
10, 2ºM, Las Palmas de Gran Canaria 35002, Spain\\\emph{}\textsuperscript{\emph{{*}}}\emph{E-mail
address}: tlazcala@yahoo.es}

\subsubsection*{Keywords}

fractals, nonlinear, stochastic, maps, complex, Fibonacci sequence
\begin{abstract}
In a recent paper by Cabrera et al. {[}\emph{Chaos, Solitons and Fractals
2021;146:110876}{]}, a linearization of DRM differences equation,
(Delayed Regulation Model), has been proposed as a scheme to explain
transfer of energy through different scales in turbulence. They claim
that this apparently simple model, by replication of Kolmogorov power
law of $k^{-5/3}$ scaling, remarks a key mechanism of behaviour for
more complex systems. Their proposal requires computation of several
products of random matrices, nevertheless they only offer an onset
of time evolution or an asymptotic approximation to all them. Also
it is suggested a fractal nature in the process of calculating the
successive characteristic polynomials of these products or the eigenvalues
of their associated self-adjoint matrices. Both questions are addressed
in this comment and are answered positively. A general formula for
the evolution in every step of mentioned stochastic linear approximation
to DRM is found as well a map from a binomial expansion of these matrices
key products to a well described fractal object.
\end{abstract}

\section{Introduction}

The Delayed Regulation Model (DRM) \cite{Mayn-1968}, a well known
workbench of population dynamics with delay, have been proposed to
replicate the transfer of energy in multi-scale cascades as described
in the Kolmogorov's -5/3 power spectrum model of turbulence \cite{Cabr-2021}.
Some of the exceptional characteristics of this equation in differences,
as the occurrence of the limit cycles and the inner chaotic dynamics
the populations near to such geometric locus suffer \cite{Poun-1980},
are attributable to the quadratic and discrete nature of the equation.
However as emergence of those behaviours is controlled with parameter
$r$ in equation
\begin{equation}
x_{g+1}=rx_{g}(1-x_{g-1})\label{eq:DRM Nstochastic}
\end{equation}
\emph{g}=0,1,2,$\dots,+\infty$, x$\in${[}0,1{]}, it is required
to reach certain value of it to qualitatively change such conducts
which range from quenching around a stable point to circulate around
a cycle \cite{Poun-1980}. And once the threshold for the existence
of a limit set has been trespassed the role of control parameter becomes
the habitual of a logistic equation \cite{Mayn-1968}. It will determine
duplications of period and transition to chaos on variable $x_{g}$.
This could be looked as the linear part of equation, --describing
the whole time-evolution near an unstable fixed point--, losing dynamic
significance with respect to the quadratic part. It would simply attest
how and at what strength population points in configuration space
are injected towards the limit set. All that because period bifurcation
subtleties are much more associated with the non-linear part.

Precisely Cabrera et al. \cite{Cabr-2021} seem to set out from the
opposite as they do not attribute a major role in the transfer of
energy through different scales in the turbulence to the nonlinear
part of map (\ref{eq:DRM Nstochastic}). The statistics of variable
$x_{g}$ in the vicinity of a fixed point is equated in their study
to the recounted energy transport. (See reference \cite{Cabr-2021}).
They assign great significance to the linear part of a modified version
of DRM in which parameter $r$ is subjected to a random variation.
The differences equation now it is read
\begin{equation}
x_{g+1}=r_{g}x_{g}(1-x_{g-1}),\label{eq:DRM Stochastic}
\end{equation}
where $r_{g}=$$b+av_{g}$, with $b>1$, $a\ge0$ real numbers, and
$v_{g}\in[0,1]$ is a random variable distributed uniformly. After
linearization of eq. (\ref{eq:DRM Stochastic}) around fixed point\footnote{Indeed Cabrera \emph{et al}. \cite{Cabr-2021} have linearized around
the other fixed point of equation (\ref{eq:DRM Nstochastic}) in plane
$\mathbb{R}^{2}$, this is the origin (0,0). Nevertheless this choice
is irrelevant for our description as the equations have the same form
and described dynamics is the same. Our option makes the affine part
of equation (\ref{eq:Lin Mod}) less significative as its mean value
is zero remarking so the homothetic part of it.} of eq. (\ref{eq:DRM Nstochastic}), --this is with $r=$$<r_{g}>$$\equiv1/(1-\alpha)$,
the average value of stochastic control parameter--, a linear affine
equation in differences for a two dimensional vectorial space $E\simeq\mathbb{R}{{}^2}$
is obtained:
\begin{equation}
\vec{X}_{g+1}=\boldsymbol{A_{g}}\vec{X}_{g}+\vec{B}_{g},\label{eq:Lin Mod}
\end{equation}
where
\[
\boldsymbol{A_{g}=}\left(\begin{array}{cc}
r_{g}(1-\alpha) & -r_{g}\alpha\\
1 & 0
\end{array}\right),\thinspace\vec{B}_{g}=\left(\begin{array}{c}
r_{g}(1-\alpha)\alpha-\alpha\\
0
\end{array}\right)
\]
which are called the homothetic matrix and the affine vector respectively.
Besides at the beginning of the series some initial conditions vector
$\vec{X}_{0}$ should be picked up preferably in a neighborhood of
$(\alpha,\alpha)^{T}$, the unstable after Hopf's bifurcation fixed
point of eq.(\ref{eq:DRM Nstochastic}). In this way its components,
and equation's evolution will start isotropically in a disc centered
in $(0,0)$, now the new coordinates of mentioned reference point
after proper shift. Finally initial conditions, near or not to zero,
will take form
\[
\vec{X}_{0}=\left(\begin{array}{c}
x_{0}-\alpha\\
y_{0}-\alpha
\end{array}\right),\quad x_{0},y_{0}\in[0,1].
\]

Iterating formula (\ref{eq:Lin Mod}) for $g\ge0$, it is easy to
see evolution of $\vec{X}_{g+1}$ via expression
\begin{equation}
\vec{X}_{g+1}=\boldsymbol{P_{g,-1}}\vec{X}_{0}+\sum_{j=0}^{g}\boldsymbol{P_{g,j}}\vec{B}_{j}\label{eq:Dyn X_n+1}
\end{equation}
 being operators $\boldsymbol{P_{g,j}}$, $j=-1,0,...,g$ the following
descending in index ordered products of matrices $\{\boldsymbol{A_{i}}\}_{i=0,1,\dots,g}$:
\[
\boldsymbol{P_{g,-1}}=\prod_{i=0}^{g}\boldsymbol{A_{g-i}}
\]
\[
\boldsymbol{P_{g,j}}=\prod_{i=0}^{g-j-1}\boldsymbol{A_{g-i}},\;j=0,1,\dots,g-1
\]
\[
\boldsymbol{P_{g,g}}=\prod_{i=0}^{-1}\boldsymbol{A_{g-i}}\equiv\boldsymbol{1}.
\]
In this case eq. (\ref{eq:Dyn X_n+1}) will give rise to a very complicate
formula with products that are not reducible to a simple one operation
of diagonalized matrices as we don't posses a common base for all
of them. That is the result of introducing a random value in every
matrix $\boldsymbol{A_{i}}$. Everything we can aspire to is obtaining
a significantly close bound to vector $\vec{X}_{g+1}$ or simulate
several times its evolution through multiple and different realizations
$\{r_{0},\dots,r_{g}\}$. Both cases, though, will require an accurate
knowledge of self-adjoint matrices $\boldsymbol{M_{g,j}}=\boldsymbol{P_{g,j}^{\dagger}P_{g,j}}$
\cite{Cabr-2021} to make a correct exposition of the evolution of
$\vec{X}_{0}$with the linear equation (\ref{eq:Lin Mod}) since they
will lavishly appear in evaluation of $||\vec{X}_{g+1}||_{2}$.

Upon reaching this state of affairs, Cabrera \emph{et al}. \cite{Cabr-2021}
made a couple of insightful observations. First of them that traces
of matrices $\boldsymbol{M_{g,j}}$ are dominant as to determine their
greatest eigenvalues, being the latters part and parcel of the bound
we are looking for. Second that traces on generation $g$, depending
on coefficients of \textbf{$\boldsymbol{A}_{\boldsymbol{i}}$}'s as
well as on stochastic parameters $r_{i}$'s, are deductible through
a recurrence relation of previous findings of the parameters over
generations $i=0,1,\dots,g-1$. (See equations 49-50 in reference
\cite{Cabr-2021}). Furthermore they put forward a step down tree
of relations among coefficients of polynomials in $\{r_{i}\}_{i=0,\dots,g-1}$
constituting such traces, and that gives account of deleting and making
of coefficients from each level to the next of series given by eq.
(\ref{eq:Dyn X_n+1}). Also they hinted for a fractal structure subjacent
to this recursion or tree but no clue about dimension was expressed.
Interesting and significative these questions as they are, were answered
with a combination of partial simulations and approximations that
rely too heavily on intuition as formal calculations were stopped
in a not so far step, $g=5$, and sets of stochastic realizations
were not fully developed \cite{Cabr-2021}.

We will address in following section the issue of writing mathematically
as far as possible matrices $\boldsymbol{M_{g,j}}$, in hope that
such formulation will give an efficient tool to answer more precisely
the former results and descriptive intuitions.

\section{Product of matrices $\boldsymbol{A_{i}}$}

The evaluation of how fast $\vec{X}_{0}$ is sinking or sourcing,
amidst noise jolts, from the fixed point $(\alpha,\alpha)^{T}$ along
generations $g=0,1,\dots$ would be quite directly estimated by mean
of euclidean distance $||\vec{X}_{g+1}||_{2}$. Unfortunately due
to the random nature of equation (\ref{eq:Lin Mod}) a simple formula
yielding a precise number it is not possible, we must be settled with
a statistical distribution of points or an upper bound to the temporal
series of vectors. To achieve this last option we'll recur to some
properties of norms as triangular inequality as well as definition
of an operator's norm, --in this case the supreme of values attained
over unit ball in $E$--, and following expression shall be gotten
\begin{equation}
\|\text{\ensuremath{\vec{X}_{g+1}}}\|_{2}\le\|\boldsymbol{P_{g,-1}}\|_{2}\|\vec{X}_{0}\|_{2}+\sum_{i=0}^{g}(\|\boldsymbol{P_{g,i}}\|_{2}\|\vec{B}_{i}\|_{2}).\label{eq:Bound X_g+1}
\end{equation}
As operator's $\boldsymbol{P}$ supreme norm definition is $\|\boldsymbol{P}\|_{2}$
$\equiv$ $\sup\frac{\|\boldsymbol{P}\vec{v}\|_{2}}{\|\vec{v}\|_{2}}$,
$\vec{0}\ne\vec{v}\in E$ \cite{Kolm-1954} all our efforts will focus
on establishing the eigenvalues of matrices $\boldsymbol{M_{g,j}}\equiv$
$\boldsymbol{P_{g,j}^{\dagger}P_{g,j}}$. To this end a general formula
for $\boldsymbol{P_{g,j}}$, $j=-1,0,\dots,g$ should be deduced as
a first step and that endeavour will start considering the definition
of every $\boldsymbol{A_{i}}$, --components of products $\boldsymbol{P_{g,j}}$--,
split in two parts one fixed and the other associated to noise and
its realizations. Every matrix $\boldsymbol{A_{i}}$ is
\[
\boldsymbol{A_{i}=}\left(\begin{array}{cc}
r_{i}(1-\alpha) & -r_{i}\alpha\\
1 & 0
\end{array}\right)=s_{i}\left(\begin{array}{cc}
1 & -\beta\\
0 & 0
\end{array}\right)+\left(\begin{array}{cc}
0 & 0\\
1 & 0
\end{array}\right)
\]
\[
\equiv s_{i}\boldsymbol{M}+\boldsymbol{N}
\]
with $s_{i}=$$r_{i}(1-\alpha)$ and $\beta=$$\alpha/(1-\alpha)$$\in(0,\infty)$.
Consequently a general ordered in descending indexes product of matrices
can be marked as 
\begin{multline*}
\boldsymbol{J_{m}}\equiv\boldsymbol{J_{m}}(s_{1},\dots,s_{m})=\prod_{i=1}^{m}(s_{i}\boldsymbol{M}+\boldsymbol{N})\\
=(s_{m}\boldsymbol{M}+\boldsymbol{N})\cdot(s_{m-1}\boldsymbol{M}+\boldsymbol{N})\cdot\ldots\cdot(s_{1}\boldsymbol{M}+\boldsymbol{N})
\end{multline*}
staying any of previous operators as $\boldsymbol{P_{g,j}}=\boldsymbol{J_{g-j}}(r_{j+1}(1-\alpha),\dots,r_{g}(1-\alpha))$,
$j=-1,0,\dots,g-1$ and $\boldsymbol{P_{g,g}}\equiv$ $\boldsymbol{1}$
$\equiv\boldsymbol{J_{0}}$ the identity matrix.

It is possible to write products $\boldsymbol{J_{m}}$ in a polynomial
form by means of a binomial-like expansion and prove that this form
contains any possible variation of two elements, $\boldsymbol{M}$
and $\boldsymbol{N}$, taken $m$times. It is a simple proof, left
to the reader as induction exercise, made easier when symbols 1 and
0 are arbitrary and respectively assigned to matrices $\boldsymbol{M}$,
$\boldsymbol{N}$. In this manner such notation allows to index products
\textbf{$\boldsymbol{Y_{m}\cdot Y_{m-1}\cdot\ldots\cdot Y_{1}}$},
--where $\boldsymbol{Y_{i}}=$ $\{\boldsymbol{M},\boldsymbol{N}\}$--,
by binary numbers of $m$ digits. Namely
\[
\boldsymbol{J_{m}}=\sum_{i=0}^{2^{m}-1}c_{i}(s_{1},\dots,s_{m})\{\boldsymbol{Y_{m}\cdot Y_{m-1}\cdot\ldots\cdot Y_{1}}\}_{i},
\]
where $c_{i}(s_{1},\dots,s_{m})$$=$$\prod_{k=1}^{m}t_{k,i}$, with
\[
t_{k,i}=\left\{ \begin{array}{c}
s_{k}\\
1
\end{array},\text{\;if k-th bit of i is\;}\begin{array}{c}
1\\
0
\end{array}\right\} ,
\]
and
\[
\boldsymbol{Y_{k}}=\left\{ \begin{array}{c}
\boldsymbol{M}\\
\boldsymbol{N}
\end{array},\text{\;if k-th bit of i is\;}\begin{array}{c}
1\\
0
\end{array}\right\} .
\]
The summation $\boldsymbol{J_{m}}$ can be grouped in $m+1$ sets
of indexes $i$ according to the number of zeroes the binary representation
each one has. A particular set will have then $\left(\begin{array}{c}
m\\
j
\end{array}\right)$ elements where now $j$ $=$ $0,1$$,\dots,$$m$ denotes the number
of zeroes of any index $i$ which belongs to it. An additional partition
of every one of these sets in four parts, for $m\ge2$, can be done
having in mind that both ends of any $m-$tuple which is a binary
representation of some $i$ has values $0$ or $1$. Writing $f11$,
$f01$, $f10$ and $f00$ for the usually non void $m-$tuples outfits
of respective type $(1,\dots,1)$, $(0,\dots,1)$, $(1,\dots,0)$
and $(0,\dots,0)$, it is obtained the following formula
\begin{equation}
\boldsymbol{J_{m}}=\sum_{j=0}^{m}\{\sum_{**\in\{11,01,10,00\}}\sum_{i\in f**}c_{i}\cdot\{\boldsymbol{Y_{m}\cdot Y_{m-1}\cdot\ldots\cdot Y_{1}}\}_{i}\}_{\#0's\;\text{in}\;i=j}.\label{eq:Binom J_m 1}
\end{equation}

\section{Sorting the products $\boldsymbol{Y_{m}\cdot Y_{m-1}\cdot\ldots\cdot Y_{1}}$}

It may seem that the previous formula (\ref{eq:Binom J_m 1}) is just
one of many multiple possible outcomes after shuffling summand in
$\boldsymbol{J_{m}}$, but really is the proper grouping of products
to reduce them to a minimum of calculations. From there on the complexity
of determining and gathering explicitly summands of $\boldsymbol{J_{m}}$
will rely on how hard is to write down coefficients $c_{i}$'s. Yet
to see that this classification of products $\boldsymbol{Y_{m}\cdot Y_{m-1}\cdot\ldots\cdot Y_{1}}$
is really optimal, we must first delve into the behaviour of products
by pairs of matrices $\boldsymbol{M}$ and $\boldsymbol{N}$.

When naming $\boldsymbol{Q}\equiv\boldsymbol{M\cdot N}$$=$$\left(\begin{array}{cc}
-\beta & 0\\
0 & 0
\end{array}\right)$ and $\boldsymbol{R}\equiv\boldsymbol{N\cdot M}$$=$$\left(\begin{array}{cc}
0 & 0\\
1 & -\beta
\end{array}\right)$ , we get a quartet of matrices, \{$\boldsymbol{M}$, $\boldsymbol{R}$,
$\boldsymbol{Q}$, $\boldsymbol{N}$\}, that forms a linearly independent
set in the four dimensional vectorial space of 2x2 matrices over reals,
and in consequence any product $\boldsymbol{Y_{m}\cdot Y_{m-1}\cdot\ldots\cdot Y_{1}}$
will be a linear combination of them with real coefficients. Besides
these four have a very nice property as they are a set closed under
matrix product which will allow to set up simple recursion formulas.
To this end we write the following table (\ref{tab:Matrix-products 1})
for the matrix products of the selected basis, it represents the multiplications
in a row times column convention.
\begin{table}
\centering{}%
\begin{tabular}{|c|c|c|c|c|}
\hline 
r{*}c & \textbf{$\boldsymbol{Q}$} & \textbf{$\boldsymbol{M}$} & \textbf{$\boldsymbol{N}$} & \textbf{$\boldsymbol{R}$}\tabularnewline
\hline 
\hline 
$\boldsymbol{Q}$ & $-\beta\boldsymbol{Q}$ & $-\beta\boldsymbol{M}$ & $\boldsymbol{0}$ & $\boldsymbol{0}$\tabularnewline
\hline 
$\boldsymbol{M}$ & $\boldsymbol{Q}$ & $\boldsymbol{M}$ & $\boldsymbol{Q}$ & $-\beta\boldsymbol{M}$\tabularnewline
\hline 
$\boldsymbol{N}$ & $-\beta\boldsymbol{N}$ & $\boldsymbol{R}$ & $\boldsymbol{0}$ & $\boldsymbol{0}$\tabularnewline
\hline 
$\boldsymbol{R}$ & $-\beta\boldsymbol{N}$ & $\boldsymbol{R}$ & $-\beta\boldsymbol{N}$ & $-\beta\boldsymbol{R}$\tabularnewline
\hline 
\end{tabular}\caption{Matrix products of the basis \{$\boldsymbol{M},\boldsymbol{R},\boldsymbol{Q},\boldsymbol{N}$\};
rows multiplies columns by the left.\label{tab:Matrix-products 1}}
\end{table}

From table (\ref{tab:Matrix-products 1}) are easily deduced by induction
the following matrix equations
\[
\boldsymbol{N}^{2}=\boldsymbol{0}\implies\boldsymbol{N}^{n}=\boldsymbol{0},\;n\ge2,
\]
\[
\boldsymbol{M}^{2}=\boldsymbol{M}\implies\boldsymbol{M}^{n}=\boldsymbol{M},\;n\ge2,
\]
\[
\boldsymbol{Q}^{2}=-\beta\boldsymbol{Q}\implies\boldsymbol{Q}^{j}=(-\beta)^{j-1}\boldsymbol{Q},\;j\ge1,
\]
\[
\boldsymbol{R}^{2}=-\beta\boldsymbol{R}\implies\boldsymbol{R}^{j}=(-\beta)^{j-1}\boldsymbol{R},\;j\ge1.
\]
This result will serve to a further reduction of every product $\boldsymbol{Y_{m}\cdot Y_{m-1}\cdot\ldots\cdot Y_{1}}$.
Each one is no other thing than an arbitrary succession of $\boldsymbol{M}$'s
or $\boldsymbol{N}$'s that can be rephrased as an alternating product
of powers of $\boldsymbol{N}$'s and $\boldsymbol{M}$'s, according
to how many neighbours of same nature remain in a row separated by
others of different nature. Being the total of multiplicands $m$
the sum of all exponents for these clusters will amount naturally
to this number. At this point a fork to classify all the products
in types is obvious due to reductions implied in the previous table
and derived subsequent equations. First alternative gives a class
of products whose result is a Zero matrix, $\boldsymbol{0}$. A particular
$\prod_{k=0}^{m}\boldsymbol{Y_{k}}$ will be in this one if any of
the exponents of all its $\boldsymbol{N}^{k}$ powers is greater than
one. The second option will comprise non null results and can be subdivided
in four other classes. As each selection of $m$ matrices now is an
alternating array of $\boldsymbol{M}$'s and $\boldsymbol{N}$'s due
to coalescence of every power of $\boldsymbol{M}$ to $\boldsymbol{M}$
itself, effectively four subtypes of arrays will be found depending
on possibilities the ends of the product allow. These are a) $\boldsymbol{M\cdot N\cdot\ldots\cdot M}$,
b) $\boldsymbol{N\cdot M\cdot\ldots\cdot M}$, c) $\boldsymbol{M\cdot N\cdot\ldots\cdot N}$
and d) $\boldsymbol{N\cdot M\cdot\ldots\cdot N}$.

Precisely the sets of binary indexes $f11$, $f01$, $f10$ and $f00$,
at every stage of $j\le m$ zeroes, represent products $\boldsymbol{Y_{m}\cdot Y_{m-1}\cdot\ldots\cdot Y_{1}}$
that will respectively give arrays of type a), b), c) and d) after
symbol redundancy is resolved. As long as, of course, no two consecutive
zeroes can be found in the inspected index belonging to $f**$. In
what follows we will understand that teams of indices $f11$, $f01$,
$f10$ and $f00$, at every level $j$, are already purged of those
$i$ whose binary representation have two or more adjacent $0$'s.

We are now in a position to calculate the product of matrices inside
$\boldsymbol{J_{m}}$ classified into classes $f11$, $f01$, $f10$
or $f00$ and levels $0\le j\le m$.

In the $j$-th stage of equation (\ref{eq:Binom J_m 1}), where $j$
denotes the number of zeroes of binary representation of indices $i$
tagging products $\{\boldsymbol{Y_{m}\cdot Y_{m-1}\cdot\ldots\cdot Y_{1}}\}_{i}$,
we will have the following results for

a) $\underbrace{\boldsymbol{M\cdot N\cdot\ldots\cdot M\cdot N}}_{j\text{pairs},j\ge1}\boldsymbol{\cdot M}$$=$$\boldsymbol{Q}^{j}\boldsymbol{\cdot M}$$=$
$(-\beta)^{j-1}\boldsymbol{Q\cdot M}$$=$ $(-\beta)^{j-1}(-\beta)\boldsymbol{M}$$=$
$(-\beta)^{j}\boldsymbol{M}$, if $j=0$ no pair $\boldsymbol{M\cdot N}$
would be present although the array being just $\boldsymbol{M}$ also
fulfill $\boldsymbol{M}=(-\beta)^{0}\boldsymbol{M}$; result is valid
then for $j\ge0$,

b) $\underbrace{\boldsymbol{N\cdot M\cdot\ldots\cdot N\cdot M}}_{j\text{pairs},j\ge1}$$=$$\boldsymbol{R}^{j}$$=$$(-\beta)^{j-1}\boldsymbol{R}$,
this time no $j=0$ case is possible since array begins with $\boldsymbol{N}$,

c) $\underbrace{\boldsymbol{M\cdot N\cdot\ldots\cdot M\cdot N}}_{j\text{pairs},j\ge1}$$=$$\boldsymbol{Q}^{j}$$=$$(-\beta)^{j-1}\boldsymbol{Q}$,
also case $j=0$ is forbidden as $\boldsymbol{N}$ ends the sequence,

d) $\underbrace{\boldsymbol{N\cdot M\cdot\ldots\cdot N\cdot M}}_{j-1\text{pairs},j\ge2}\boldsymbol{\cdot N}$$=$$\boldsymbol{R}^{j-1}\boldsymbol{\cdot N}$$=$
$(-\beta)^{j-2}\boldsymbol{R\cdot N}$$=$ $(-\beta)^{j-2}(-\beta)\boldsymbol{N}$$=$
$(-\beta)^{j-1}\boldsymbol{N}$, $j\ge2$ is necessary as array starts
and ends by $\boldsymbol{N}$.\footnote{Formula is also valid for $j=1$ as $\boldsymbol{R}^{0}\boldsymbol{\cdot N}=(-\beta)^{0}\boldsymbol{N}=\boldsymbol{N}$,
and is required to formally extent this treatment to $m=1$, the trivial
case for $\boldsymbol{J_{m}}$. In this situation only a) y d) sets,
and $j=0,1$ levels, are present in eq. (\ref{eq:Binom J_m 1}). In
the nontrivial cases $m\ge2$, when $j=0,1$, d) set is empty.}

\subsection{Void $f11$, $f01$, $f10$, $f00$ populations}

Except for those few cases in which the use of binomial coefficients
implies a negative factorial, --a zero result then--, the original
populations of sets $f11$, $f01$, $f10$, $f00$, at level $j$,
contain respectively $\left(\begin{array}{c}
m-2\\
j
\end{array}\right)$, $\left(\begin{array}{c}
m-2\\
j-1
\end{array}\right)$, $\left(\begin{array}{c}
m-2\\
j-1
\end{array}\right)$, and $\left(\begin{array}{c}
m-2\\
j-2
\end{array}\right)$ elements, ($j\ge2$, $m-j\ge2$). Nevertheless this quantities will
be depleted in a fractal look as greater values of $m$ and $j$ are
considered due to the ruling out of binary sequences with adjacent
zeros. Not being interested by now on the geometry of such decimation,
which is a problem to pose in next sections, we just will count in
next paragraphs those stages with too many zeros as to have no population
at all in sets $f**$ since their wiping out makes the formula (\ref{eq:Binom J_m 1})
clearer to write.

For example, we observe that if $2j>m$ the outfits $f01$ and $f10$
would never give rise to any non null product of type b) or c) since
the number of matrices $\boldsymbol{N}$ is greater than $\boldsymbol{M}$'s
and $j$ pairs of $\boldsymbol{M\cdot N}$ products, or $\boldsymbol{N\cdot M}$,
cannot be formed. This will lighten up the second summation in equation
(\ref{eq:Binom J_m 1}) from a particular index of all possible $j\in\{0,1,\dots,m\}.$
With the intent of figuring it out we must discriminate two possible
situations: $m$ is even, ($m=2[m/2]$), or $m$ is odd, ($m=2[m/2]+1$),
so elimination condition it will be read now
\[
j>\left[\frac{m}{2}\right]+\left\{ \begin{array}{cc}
0 & \text{if \ensuremath{m} is even}\\
\frac{1}{2} & \text{if \ensuremath{m} is odd}
\end{array}\right\} ,
\]
though as $j$ must be an entire number the latter simplifies to $j>[m/2]$.

Also populations $f11$ and $f00$, --which after simplification
drive to corresponding product types a) and d)--, will be restricted
based on the number of zeroes, $j$, in their binary representations.

In case $f00\rightarrow$d) the number of zeroes minus one, $j-1$,
which represents the number of matrices $\boldsymbol{N}$ paired to
the right with at least one matrix $\boldsymbol{M}$, must be less
or equal to the number of ones, $m-j$, to exist. This is, case d)
is obliterated from our accountancy if $j-1>m-j$, and again a condition
with multiples of $j$ it is not suitable for use in equation (\ref{eq:Binom J_m 1}),
so $2j>m+1$ will be expressed as
\[
j>\left[\frac{m}{2}\right]+\left\{ \begin{array}{cc}
0 & \text{if \ensuremath{m} is even}\\
1 & \text{if \ensuremath{m} is odd}
\end{array}\right\} .
\]

In case $f11\rightarrow$a) the number of $\boldsymbol{M}$'s minus
one must be at least equal to the number of $\boldsymbol{N}$'s to
exist, although can be greater of course. This is $m-j-1\ge j$ and
consequently case a) cannot be possible if $m-1<2j$, or what is the
same
\[
j>\left[\frac{m}{2}\right]+\left\{ \begin{array}{cc}
-1 & \text{if \ensuremath{m} is even}\\
0 & \text{if \ensuremath{m} is odd}
\end{array}\right\} .
\]

The first summation in eq. (\ref{eq:Binom J_m 1}) will be reduced
from $\sum_{j=0}^{m}$ to $\sum_{j=0}^{\left[\frac{m}{2}\right]+1}$
for being null summands $j\ge[m/2]+2$, (cases a), b), c) or d) are
not possible from elements in any $f**$). From the remaining summands
we have already discussed that indices $2\le j\le[m/2]-1$ will raise
to cases a), b), c) and d), yet $j=0$ only to a) and $j=1$ to a),
b), c) cases but not to d) one.\footnote{It is an exception $m=2$ with $j=1$ as a) case is (1,1) and contains
no zeroes.} And just as at the beginning of the count all cases are not present
and it is required for summands fulfill $j\ge2$ in order to contain
the four cases, the end of summation is not abrupt at $j=[m/2]+1$
either. Terms with $j=$ $[m/2]$ or $[m/2]+1$ not always will have
all four, yet depends on parity of $m$ to know which ones survive.

If $m$ is even and:
\begin{itemize}
\item $j=\left[\frac{m}{2}\right]$, a) it does not ride out, but b), c)
and d) do,
\item $j=\left[\frac{m}{2}\right]+1$, neither of a), b), c) or d) rides
out.
\end{itemize}
If $m$ is odd and:
\begin{itemize}
\item $j=\left[\frac{m}{2}\right]$, all cases, (a, b, c, d), survive,
\item $j=\left[\frac{m}{2}\right]+1$, a), b), and c) they don't ride out
the cut but d) does.
\end{itemize}
This is all there is to consider in relationship to which product
sequences $\boldsymbol{Y_{m}\cdot Y_{m-1}\cdot\ldots\cdot Y_{1}}$
disappear due to nilpotency of $\boldsymbol{N}$ and the canonical
types the survivors fall into. A closure in the characterization of
\textbf{$\boldsymbol{J}_{\boldsymbol{m}}$} summands requires to write
properly remaining coefficients $c_{i}(s_{1},\dots,s_{m})$ in function
of indices $i$ and $j$ and families $f00$, $f01$, $f10$, and
$f11$ they belong to.

\section{Coefficients $c_{i}(s_{1},\dots,s_{m})$}

We saw that $c_{i}$'s are productories with $m$ terms $t_{k,i}$,
($k=1,\dots,m$), the latter being one two choices: the random variable
$s_{k}$ or $1$ depending upon the binary representation of $i$$=$$0,\dots,2^{m}-1$.
At the moment the best description for them since all numbers $i$
of $m$ binary digits where included, nevertheless the suitable $c_{i}$'s
for the equation (\ref{eq:Binom J_m 1}) are obtained after two selective
processes, a sort and a purging one. The first is a grouping of summands
according to the number of zeroes, $j$, in the binary representation
of $i$; the second is a overriding of those same sequences when two
or more adjacent zeroes exist in them. As such it looks as convenient
a change of notation in $c_{i}$'s to reflect all this and make an
indexed use of them easier.

To this end let us define as auxiliary functions the products of consecutive
variables $x_{s}$, $s\in\{1,\dots,m\}$, specifically they shall
be written in this guise $\pi_{l}^{n}(x_{l+1},\dots,x_{l+n})$$=$$\prod_{k=1}^{n}x_{l+k}$,
for $n\ge1$, or $\pi_{l}^{0}$$\equiv$$1$, for $n=0$.

Also the binary representation of $i$ will be additionally tagged
again. Originally to a unique index $i$ for each product of matrices
was added a second index, $j$, due to operative reasons as shown
in formula (\ref{eq:Binom J_m 1}), however outfits $f**$ need also
a more descriptive and possibly more efficient third system, since
it will facilitate a systematic writing of those valid $c_{i}$ coefficients.

Two additional marked positions, $m+1$ and $0$, will be added to
left and right of a binary $m$-tuple, $i$$\in$$\{0,\dots,2^{m}-1\}$,
with the purpose of accounting for loci of zeros and clustering of
ones. As it is highlighted the number of zeros of each sequence, --or
binary representation--, with number $j$, they will numbered with
indices $k=1,\dots,j$ counting them from right to left, and their
locations logged with indices $l_{k}$. In this way $1\le$ $l_{1}<\dots<l_{j}$
$\le m$, and $l_{0}$$\text{\ensuremath{\equiv}}0$, $l_{j+1}$$\equiv m+1$
always, what allows to write the number of ones between consecutive,
and not necessarily adjacent, zeros as $n_{k}=$$l_{k+1}-l_{k}-1$,
with $0\le$$k$$\le j$. Obviously the total number of $1$'s in
the representation of $i$ is obtained without ambiguity, \emph{i.e.}
$m-j$$=$$\sum_{k=0}^{j}n_{k}$, and with the zeroes as milestones,
neither their positions are subjected to confusion. Now, --having
in mind it has been abridged every $l_{k}(i)$ to $l_{k}$--, we
can rewrite all coefficients as
\[
c_{i}(s_{1},\dots,s_{m})=\prod_{k=0}^{j}\pi_{l_{k}}^{n_{k}}(s_{l_{k}+1},\dots,s_{l_{k}+n_{k}}).
\]
This wouldn't be a great change of notation but for we have previously
rejected use of all $c_{i}$'s associated with null products of matrices,
(this is $i$'s with adjacent $0$'s). Since from now on $n_{k}\ge1$,
for $k=1,2,\dots,j-1$, the latter will be the most compact form of
writing such coefficients for families $f11$, $f01$, $f10$ and
$f00$, since it will involve a minimal use of functions $\text{\ensuremath{\pi_{l_{k}}^{n_{k}}}}$
containing random variables \{$s_{1},...,s_{m}$\}. Besides any padding
$1$'s is excluded of $\prod_{k=1}^{j-1}$ unlike when use of symbols
$t_{k,i}$ in $c_{i}$ was habitual.

The described set of new indexes based on numbers $m$, $j$, and
grouping of ones $\{l_{k}\}$, as well as aforementioned conditions
on $n_{l_{k}}$ will be denoted with symbol
\[
\mathcal{P}_{m,j}=\left\{ (l_{j},\dots,l_{1})|l_{k}\in\{1,\dots,m\};\{n_{k}\ge1\}_{k=1,\dots,j-1};\sum_{k=0}^{j}n_{k}=m-j\right\} .
\]

\section{The binomial expansion of $\boldsymbol{J_{m}}$}

We rephrase eq. (\ref{eq:Binom J_m 1}) for each level $j$$\in$\{$0,1,\dots,[m/2]+1$\}
as follows, each element $i$ from any of four outfits $f**,_{j}$
will be substituted by the corresponding $j-$tuple of $\mathcal{\mathcal{P}}_{m,j}$
as counting index in the inner summations. In this way
\[
a,b,c,d)\sum_{{i\in f**\atop \#0's\;\text{in}\;i=j}}c_{i}\rightarrow\left\{ \sum_{{\boldsymbol{l}\in\mathcal{P}_{m,j}\atop \mathcal{C}**(n_{0},n_{j})}}\prod_{k=0}^{j}\pi_{l_{k}}^{n_{k}}(s_{l_{k}+1},\dots,s_{l_{k}+n_{k}})\right\} \equiv p_{**,j},
\]

where if a) $**$$=$$11$ then $\mathcal{C}11(n_{0},n_{j})$$=$
$\{n_{0},n_{j}\ge1\}$, if b) $**$$=$$01$ then $\mathcal{C}01(n_{0},n_{j})$$=$
$\{n_{0}\ge1,n_{j}=0\}$, if c) $**=10$ then $\mathcal{C}10(n_{0},n_{j})$$=$
$\{n_{0}=0,n_{j}\ge1\}$ and if d) $**$$=$$00$ then $\mathcal{C}00(n_{0},n_{j})$$=$
$\{n_{0}=n_{j}=0\}$.

Finally we obtain the coveted expression for equation
\[
\boldsymbol{J_{m}}=\sum_{j=0}^{[\frac{m}{2}]+1}\left\{ p_{11,j}(-\beta)^{j}\boldsymbol{M}+p_{01,j}(-\beta)^{j-1}\boldsymbol{R}+p_{10,j}(-\beta)^{j-1}\boldsymbol{Q}+p_{00,j}(-\beta)^{j-1}\boldsymbol{N}\right\} ,
\]
with $p_{01,j}$$=$ $p_{10,j}$$\equiv$ $0$, for $j=0$, and $p_{00,j}$$\equiv$$0$
for $j=0$ or $j=1$. Also is possible to write again this formula
by extracting common matricial factors out of summations, we will
obtain then
\begin{equation}
\boldsymbol{J_{m}}(\vec{s}_{m})=g_{12}(\vec{s}_{m})\boldsymbol{M}+g_{22}(\vec{s}_{m})\boldsymbol{R}+g_{11}(\vec{s}_{m})\boldsymbol{Q}+g_{21}(\vec{s}_{m})\boldsymbol{N},\label{eq:Binom J_m 2}
\end{equation}
where $\vec{s}_{m}$ stands for $(s_{1},\dots,s_{m})$. Obviously
functions $g_{xy}$ will be
\[
g_{12}(\vec{s}_{m})=\sum_{j=0}^{[\frac{m}{2}]+1}p_{11,j}(-\beta)^{j},
\]
\[
g_{22}(\vec{s}_{m})=\sum_{j=1}^{[\frac{m}{2}]+1}p_{01,j}(-\beta)^{j-1},
\]
\[
g_{11}(\vec{s}_{m})=\sum_{j=1}^{[\frac{m}{2}]+1}p_{10,j}(-\beta)^{j-1},
\]
\[
g_{21}(\vec{s}_{m})=\sum_{j=2}^{[\frac{m}{2}]+1}p_{00,j}(-\beta)^{j-1}.
\]

\section{Eigenvalues of $\boldsymbol{M_{g,j}}$ and norm of $\boldsymbol{P_{g,j}}$}

Once the structure of $\boldsymbol{J_{m}}$ have been clarified it
is straightforward to get $\boldsymbol{J_{m}^{\dagger}J_{m}}$. We
skip details here as the result pops up after easy and laborious calculations
facilitated by inspection of table (\ref{tab:Matrix-products 2}).
\begin{table}
\centering{}%
\begin{tabular}{|c|c|c|c|c|}
\hline 
r{*}c & \textbf{$\boldsymbol{Q}$} & \textbf{$\boldsymbol{M}$} & \textbf{$\boldsymbol{N}$} & \textbf{$\boldsymbol{R}$}\tabularnewline
\hline 
\hline 
$\boldsymbol{Q^{\dagger}}$ & $-\beta\boldsymbol{Q}$ & $-\beta\boldsymbol{M}$ & $\boldsymbol{0}$ & $\boldsymbol{0}$\tabularnewline
\hline 
$\boldsymbol{M^{\dagger}}$ & $-\beta\boldsymbol{M^{\dagger}}$ & $\boldsymbol{M}-\beta\boldsymbol{R}$ & $\boldsymbol{0}$ & $\boldsymbol{0}$\tabularnewline
\hline 
$\boldsymbol{N^{\dagger}}$ & $\boldsymbol{0}$ & $\boldsymbol{0}$ & $-\frac{1}{\beta}\boldsymbol{Q}$ & $\boldsymbol{M}$\tabularnewline
\hline 
$\boldsymbol{R^{\dagger}}$ & $\boldsymbol{0}$ & $\boldsymbol{0}$ & $\boldsymbol{M^{\dagger}}$ & $\boldsymbol{M}-\beta\boldsymbol{R}$\tabularnewline
\hline 
\end{tabular}\caption{Products of transposes of \{$\boldsymbol{M},\boldsymbol{R},\boldsymbol{Q},\boldsymbol{N}$\}
by original ones; rows multiplies columns by the left.\label{tab:Matrix-products 2}}
\end{table}
 The outcome it is read as
\begin{equation}
\boldsymbol{J_{m}^{\dagger}J_{m}}=(-\beta g_{11}^{2}-\frac{1}{\beta}g_{21}^{2})\boldsymbol{Q}+(-\beta g_{11}g_{12}+g_{22}g_{21})(\boldsymbol{M^{\dagger}}+\boldsymbol{M})+(g_{12}^{2}+g_{22}^{2})(\boldsymbol{M}-\beta\boldsymbol{R}).\label{eq:Bin M_g,j 1}
\end{equation}
However as matrices and coefficients in previous formula both show
a dependence in parameter $\beta$, it is tidier and more efficient
operationally to split this formula in a scalar part depending on
$\beta$ and a vectorial one not doing so. We will resort to use three
additional matrices to make this possible. They are $\boldsymbol{K}=$
$\left(\begin{array}{cc}
1 & 0\\
0 & 0
\end{array}\right)$, $\boldsymbol{L}=$ $\left(\begin{array}{cc}
0 & 1\\
1 & 0
\end{array}\right)$, $\boldsymbol{S}=$ $\left(\begin{array}{cc}
0 & 0\\
0 & 1
\end{array}\right)$, and with them the matrices involved in $\boldsymbol{J_{m}^{\dagger}J_{m}}$
are written as $\boldsymbol{Q}=$ $-\beta\boldsymbol{K}$, $\boldsymbol{M^{\dagger}}+\boldsymbol{M}=$
$2\boldsymbol{K}-\beta\boldsymbol{L}$ and $\boldsymbol{M}-\beta\boldsymbol{R}=$
$\boldsymbol{K}-\beta\boldsymbol{L}+\beta^{2}\boldsymbol{S}$. Introducing
these identities in the equation, expanding parenthesis and rearranging
all terms in function of matrices $\boldsymbol{K}$, $\boldsymbol{L}$
and $\boldsymbol{S}$ a new version of eq. (\ref{eq:Bin M_g,j 1})
is achieved
\begin{multline}
\boldsymbol{J_{m}^{\dagger}J_{m}}=\left\{ (\beta g_{11}-g_{12})^{2}+(g_{21}+g_{22})^{2}\right\} \boldsymbol{K}\\
+\beta\left\{ g_{12}(\beta g_{11}-g_{12})-g_{22}(g_{21}+g_{22})\right\} \boldsymbol{L}\\
+\beta^{2}\left\{ g_{12}^{2}+g_{22}^{2}\right\} \boldsymbol{S}.\label{eq:Bin M_g,j 2}
\end{multline}
That symmetric matrix is shorten to $\boldsymbol{J_{m}^{\dagger}J_{m}}$
$=$ $\left(\begin{array}{cc}
h_{11} & h_{12}\\
h_{21} & h_{22}
\end{array}\right)_{m}$ to point out the functional look of its two real and positive eigenvalues
\begin{equation}
\lambda_{m,\pm}=\left(\frac{h_{11}+h_{22}\pm\sqrt{(h_{11}-h_{22})^{2}+4h_{12}h_{21}}}{2}\right)_{m},\label{eq:Eigenvalues M_g,j}
\end{equation}
of which the greater is, by the definition given in section 2, the
square of operator's, $\boldsymbol{J_{m}}$, norm, (\emph{i.e.} $||\boldsymbol{J_{m}}||_{2}^{2}=\lambda_{m,+}(\vec{s}_{m})$
\cite{Kolm-1954}).

Finally all requirements to describe dynamics of equation (\ref{eq:Dyn X_n+1})
end here, as formula (\ref{eq:Binom J_m 2}) provides the elements
needed to follow its evolution in time. However we have continued
a little further in the search of an mathematical expression for $\boldsymbol{M_{g,j}}$
as it allows for an unidimensional picture of the linearized DRM difference
system. These matrices, we saw, are built in a natural way as self-adjoint
operators derived from corresponding \textbf{$\boldsymbol{P_{g,j}}$}'s,
though. And in section 2 was shown that $\boldsymbol{P_{g,j}}=\boldsymbol{J_{g-j}}(\vec{s}_{g-j})$,
with $s_{k}$$=$$r_{j+k}(1-\alpha)$, $j=-1,0,\dots,g-1$, and $k=1,\dots,g-j$,
so accordingly to equation (\ref{eq:Bound X_g+1}) the former eigenvalues
in eq. (\ref{eq:Eigenvalues M_g,j}), considered as functions of random
variables, furnish all is needed for making up every one of the norms
required to bound closely $||\vec{X}_{g+1}||_{2}$, once is given
a realization \{$r_{0},\dots,r_{g}$\}.

\section{Subjacent fractal distribution of matrices' products}

The question now is to ascertain how many matrices' products $\prod_{k=1}^{m}\boldsymbol{Y_{k}}$
among those initial $2^{m}$ have survived after being purged by nilpotence
of $\boldsymbol{N}$, inasmuch as its answer will make easier and
efficient the writing of partition sets $\mathcal{P}_{m,j}$ and consequently
that of functions $g_{\{1,2\}\{1,2\}}(\vec{s}_{m})$ in equation (\ref{eq:Binom J_m 2}).
To take one step back it is needed then, and all products null or
ending in classes a) to d) must be again considered. Explicitly the
index $i$ attributed to each one will be written down $i=$ $\sum_{k=0}^{m-1}i_{k}2^{k}$,
with $i_{k}\in$ $\{0,1\}$, or in $m-$tuple form $(i_{m-1},\dots,i_{1},i_{0})$,
this latter is an ever growing set of indices, --increasing with
the number of generations $m$ tried--, that is hard to apprehend
in a geometric picture. Nevertheless it is always possible mapping
all indexes to the finite interval $[0,1]$ $\subset\mathbb{R}$ by
mean of inversion and study the distribution of survivors. Henceforth
at every generation the interval $[0,1]$ will be divided in $2^{m}$
equal subintervals and they will be numbered following binary notation
from $(0,\dots,0)$ to $(1,\dots,1)$, these will be associated to
indices $i$'s as it was done before but this time reading the $m-$tuples
in reverse bit order. Every index with this idea in mind will be paired
with a subinterval of extent $1/2^{m}$ whose left end in a fractional
binary representation is the new $m-$tuple. This is, if a matrix
product was indexed with an $i$ $\in$ \{$0,\dots,2^{m}-1$\}, as
just described above now it will be assigned to subinterval starting
at boundary post $I_{i}=$ $\sum_{k=1}^{m}i_{k-1}2^{-k}$. Such indexation
has a tremendous advantage, since as generations run they are drawn
in a stack and on it each subinterval at layer $m-1$ is split in
two which will be put just beneath their parent in the next layer,
also a consecutive numbering is hold among all members of the new
generation $m$. Besides this method, or image, reflects exactly how
products constituting $\boldsymbol{J_{m-1}}$ will sire those new
elements of $\boldsymbol{J_{m}}$, and how to neglect those they are
null and record the survivors with a huge economy of means that avoids
effectively count all the $2^{m}$ products.

The procedure as told contains all elements necessary for registering
in a descending tree of decisions all cases. A simple two steps system
is all what is needed to depict correctly the geometry of every layer
as well as the limit set. We observe when $m=1$ a simple division
in two of segment $[0,1]$, subsegments are numbered $0$ and $1$.
The next layer, $m=2$, split each one of the previous in two and
results are tagged as $00$, $01$, $10$ and $11$. This pattern
is crucial since every two layers, from $m$ even to $m+2$, every
subsegment will be split in four and these same mentioned tags will
be added to the sequence of ones and zeroes each interval already
has assigned. And this is the first hint of a fractal structure in
the limit set $m$$\rightarrow$$\infty$.

The Cantor set is the result of deleting indefinitely the middle third
at every turn the remains of interval $[0,1]$ which is that was started
with. It has a fractal dimension of $0.631$ and it is the classical
example of self-similarity when these objects are introduced. We face
here something lookalike though different. As we only allow chains
of symbols with no consecutive zeroes one quarter is wiped out of
present segments every two steps of duplicating segments and the process
is iterated also indefinitely. Anyone can argue this is a bad and
non-symmetric copy of Cantor's set, but no argument against its fractality
can be issued. Nevertheless there is still more, the three segments
$01$, $10$ and $11$ alive at step $m$ once duplicated they become
in $010$, $011,$$100$, $101$, $110$ and $111$ at step $m+1$
and five of them survive for an already explained further pruning
at $m+2$. As deduced from inspection after adding $0$'s or $1$'s
to the binary fractional numbers which divide interval $[0,1]$ at
each step, no other patterns of elimination of subintervals are visible.
Always at each layer there are groupings of two or three neighboring
intervals of scale $2^{-m}$ and after a splitting the first type
of groupings sires three subintervals of scale $2^{-m-1}$ and the
second class five. These are then all the rules for characterizing
the fractal we observe.

While these guidelines allow to describe a fractal, $\mathcal{F}$,
in an iterative mode we simply can in a first instance to use the
easy original rule of no ``adjacent zeros'' to examine and count
filled boxes, --or not neglected intervals--, along a few steps
$m$ with intend of esteeming a fractal dimension which tells how
many surviving intervals are found at each scale of division $\delta$.
The implicit model to account for content of a fractal would be $\mathcal{M}_{\delta}(\mathcal{\mathcal{F}})$
$\sim$ $\text{\ensuremath{\mathcal{C}}\ensuremath{\delta}}^{-s}$,
where $\mathcal{M}_{\delta}$ is a function which answer how much
\emph{matter} of the object $\mathcal{F}$ is found at scale $\delta$
and $s$ is the fractal dimension of it. In our description of $\mathcal{F}$
we have counted remaining intervals of length $\delta$ $=$ $1/2^{m}$
after $m$ layers of pruning, that is our $\mathcal{M}_{\delta}$
then. In this way of things a formula to work out dimension $s$ is
\cite{Falc-1990}
\[
s=\lim_{\delta\rightarrow0}\frac{\ln\mathcal{M}_{\delta}(\mathcal{F})}{-\ln\delta}.
\]

\begin{table}
\begin{centering}
\begin{tabular}{|c|c|c|c|c|c|c|c|c|c|c|}
\hline 
$\ln\mathcal{M}_{\text{\ensuremath{\delta}}}$ & 1 & 3 & 5 & 8 & 13 & 21 & 34 & 55 & 89 & 144\tabularnewline
\hline 
\hline 
$m$ & 1 & 2 & 3 & 4 & 5 & 6 & 7 & 8 & 9 & 10\tabularnewline
\hline 
\end{tabular}
\par\end{centering}
\caption{Number of boxes $\mathcal{M}_{\delta}$ at division scale $\delta$
$=$ $2^{-m}$ \emph{vs.} number of generations $m$ in the splitting
process of interval $[0,1]$.\label{tab:Number of boxes}}
\end{table}
In table (\ref{tab:Number of boxes}) are data needed to justify the
following regression results of formula $\ln\mathcal{\mathcal{M}}_{\delta}\sim$
$s(m\ln2)$ $+$ $\ln\mathcal{C}$: correlation coefficient, $\text{\ensuremath{\rho=}}$
$0.9999229$, constant term, $\ln\mathcal{C}=$ $0.173$ $\pm$ $0.013$,
slope $s=$ $0.691$ $\pm$ $0.003$. This ends our problem of counting
matrices in $\boldsymbol{J_{m}}$.

\subsection{The Fibonacci Sequence}

However a final remark must be added on the light of explained ``2-3
grouping to 3-5 survivors'' rule used for reproducing non void subintervals
and a remarkable fact, shown in table (\ref{tab:Number of boxes}),
which is that the number of boxes recorded follows a Fibonacci sequence.
With the final purpose of giving an homogeneous treatment to both
kinds of groupings we start a description of their spawning at the
level $m=3$ where only the intervals $010$, $011$, --both contiguous--,
and $101$, $110$, $111$, --also neighbors--, have survived to
the purge. This level is chosen as is the first time the mentioned
groups of two or three subintervals are present together and henceforth
they will be appearing persistently in the ongoing levels. In contrast
when $m=0$ is the whole interval $[0,1]$ which stands, level $m=1$
have a unique grouping of two subintervals, and $m=2$ consists of
an array of three adjacent surviving intervals of length $\delta$
$=$ $0.25$ each.

It is easy to see from our rule of non adjacent zeros that the set
of two splits in three elements with endings already known and the
set of three produces five intervals. A surprise comes now as those
five are not contiguous ones but they form a group of two and a group
of three whose endings, --last three ciphers--, are equal to those
recently written so that the process of division repeats indefinitely
and only groupings of two or three elements contribute to the total
of surviving intervals. Hence a formula can be written, be $m\ge3$,
$D_{m}$ the number of groups of two subintervals and $T_{m}$ the
number of groups of three elements so grand total will be
\[
\#Boxes_{m}=2\times D_{m}+3\times T_{m}
\]
and as a consequence of explained reproduction scheme on next level,
$m+1$, they will result in
\[
\#Boxes_{m+1}=2\times T_{m}+3\times(D_{m}+T_{m}).
\]
In this way coefficients fulfill a recurrence ratio
\[
D_{3}=0,T_{3}=1,
\]
\[
\left\{ {D_{m+1}=T_{m}\atop T_{m+1}=T_{m}+D_{m}}\right\} ,m\ge3
\]
which is no other than the recurrence which defines Fibonacci sequence
though with index shifted by two. On the other hand we have just shown
by a hand calculation that the number of non discarded subintervals
at level $m$ follows also a Fibonacci sequence shifted by minus two
for $0\le$ $m$ $\le10$. If this pattern continued indefinitely
the formula would be
\begin{equation}
F_{m+2}=2\times F_{m-2}+3\times F_{m-1},\,m\ge3,\label{eq:QFib rec 1}
\end{equation}
being $F_{n}$, $n\ge0$ the $n-$th term of Fibonacci sequence.

In conclusion if we can prove such formula the problem of counting
filled boxes which defines the content at each level of fractal $\mathcal{F}_{m}\rightarrow\mathcal{F}$
and of working out its dimension would be solved.

\subsubsection{A recurrence quadratic formula}

Starting with the solutions to quadratic equation $x^{2}$$-$$x$$-$$1$$=$$0$
two independent sequences of powers of them can be built which satisfy
the Fibonacci's recurrence $x_{\pm}^{n+1}$$=$$x_{\pm}^{n}$$+$$x_{\pm}^{n-1}$,
$n\ge0$. Neither of both will be sequences of entire numbers, nevertheless.
Though as they are $x_{+}$$=$$(1+\sqrt{5})/2$ $\equiv\phi$ and
$x_{-}$$=$$(1-\sqrt{5})/2$ $\equiv\psi$ some linear combination
of powers is expected to yield entire numbers at every step, and indeed
so it is. The well known class of linear combinations, (Binet's formula)
\cite{Hons-1985},
\[
F_{n}=\frac{\phi^{n}-\psi^{n}}{\phi-\psi},\,n\ge0,
\]
fulfills the recurrence
\[
F_{n+1}=F_{n}+F_{n-1},\,n\ge1,
\]
and as $F_{0}$$=$$0$ and $F_{1}$$=$$1$ results in a sequence
of natural numbers.

Once the general term is presented in this form it is quite easy to
prove \cite{Hons-1985,Basi-1963}
\begin{equation}
F_{l}F_{n}+F_{l+1}F_{n+1}=F_{l+n+1},\,l,n\ge0.\label{eq:QFib rec 2}
\end{equation}
Let's see it. On one side
\begin{multline*}
F_{l}F_{n}+F_{l+1}F_{n+1}=\\
\frac{(\phi^{l+n}-\psi^{l}\phi^{n}-\phi^{l}\psi^{n}+\psi^{l+n})+(\phi^{l+n+2}-\psi^{l+1}\phi^{n+1}-\phi^{l+1}\psi^{n+1}+\psi^{l+n+2})}{(\phi-\psi)^{2}}\\
=\frac{\phi^{l+n}(1+\phi^{2})-(\psi^{l}\phi^{n}+\phi^{l}\psi^{n})(1+\psi\phi)+\psi^{l+n}(1+\psi^{2})}{(\phi-\psi)^{2}}\\
=\frac{\phi^{l+n}(1+\phi^{2})+\psi^{l+n}(1+\psi^{2})}{(\phi-\psi)^{2}},
\end{multline*}
since $1+\psi\phi$$=$$0$, and we hold apart this in mind for a
moment. On the other hand if we take into account that $\phi$$(1+\psi^{2})$
$=$ $(\phi-\psi)$ $=$ $-\psi$$(1+\phi^{2})$, we will get 
\begin{align*}
F_{l+n+1} & =\frac{(\phi-\psi)(\phi^{l+n+1}-\psi^{l+n+1})}{(\phi-\psi)^{2}}\\
= & \frac{-\psi\phi^{l+n+1}(1+\phi^{2})-\phi\psi^{l+n+1}(1+\psi^{2})}{(\phi-\psi)^{2}}\\
= & \frac{\phi^{l+n}(1+\phi^{2})+\psi^{l+n}(1+\psi^{2})}{(\phi-\psi)^{2}}.
\end{align*}
As both expressions are the same one, identity (\ref{eq:QFib rec 2})
is proven.\footnote{Though not so well known, formula (\ref{eq:QFib rec 2}) has also
been baptized by practitioners in the field, this time as Honsberger
identity \cite{Hons-1985,Basi-1963}. Nevertheless this last one can
be also be derived straightforwardly from another more familiar one,
d'Ocagne identity. Although negative indices, not only positives,
must be taken into consideration.} Recalling that $F_{3}=2$ and $F_{4}=3$ we can substitute $l=3$
and $n=m-2$ in formula (\ref{eq:QFib rec 2}) to obtain formula (\ref{eq:QFib rec 1}).

\subsection{Fractal dimension}

Therefore it has been proved that the number of non-empty boxes of
length $\delta$ $=$ $1/2^{m}$ in $\mathcal{F}_{m}$, $m\ge3$ is
$F_{m+2}$ according to equation (\ref{eq:QFib rec 1}) secured by
the spawn of boxes present in step $m-1$. (In fact, as a digression,
we are compelled to acknowledge validity of formula (\ref{eq:QFib rec 1})
also for $0\le$ $m$$<3$ as a direct calculation shows.\footnote{It is needed then to extent the use of Fibonacci sequence to negative
subindices. In such cases $F_{-n}=(-1)^{n+1}F_{n}$.}) So we are in a position to work out precisely the already mentioned
fractal dimension. Now taking into account that $\ln\mathcal{M}_{\delta}(\mathcal{F})$
$=$ $\ln F_{m+2}$ a seamless recast of such quantity as $s(m\ln2)$
$+$ $\ln\mathcal{C}$, embodies like
\begin{multline*}
\ln F_{m+2}=\ln\phi^{m+2}+\ln(1-\left(\frac{\psi}{\phi}\right)^{m+2})-\ln(\phi-\psi)\\
=(m\ln2)\left(\frac{\ln\phi+\frac{1}{m}\ln(1-(\psi/\phi)^{m+2})}{\ln2}\right)+\ln\left(\frac{1+\phi}{\phi-\psi}\right).
\end{multline*}
And as it also happens that $|\psi/\phi|$ $=$ $|\frac{1-\sqrt{5}}{1+\sqrt{5}}|$
$<1$ results in $\lim_{m\rightarrow\infty}(1-(\psi/\phi)^{m+2})$
$=$ $1$, hence as expected a simple asymptotic formula is obtained
\begin{equation}
\mathcal{M}_{\delta}(\mathcal{F})=F_{m+2}\sim\mathcal{C}\delta^{-s}\label{eq:Fract Fib Measure}
\end{equation}
for $\delta\rightarrow0$, (\emph{i.e.} $m\rightarrow\infty$), which
yield results $s=$ $\frac{\ln\phi}{\ln2}$ $=$ $0.694241\dots$
and $\mathcal{C}=$ $\frac{1+\phi}{\phi-\psi}$ $=$ $1.17082\dots$
as fractal dimension and content respectively.

We would like to finish our exposition of current problem with a remark
highlighting how much information formula (\ref{eq:Fract Fib Measure})
conveys. Values of $m$ not near of being suspicious for justifying
formula as $m=$$0,1,2,3$ gives respective countings of $1.171$,
$1.894$, $3.065$ and $4.960$ instead of correct answers $1$, $2$,
$3$ and $5$. Really not a very bad closeness, although responses
improve to give at $m=20$ a number of boxes of $17711.00001$ being
$F_{22}=$$17711$.

\section{Conclusions}

In this work has been shown how to proceed from the linearization
of a canonical class of maps with memory describing population dynamics,
--DRM and its manipulations in eqs. (\ref{eq:DRM Nstochastic}),
(\ref{eq:DRM Stochastic}) and (\ref{eq:Lin Mod})--, to an expression
bounding its evolution in eq. (\ref{eq:Bound X_g+1}) in which the
key point to focus attention on was the product of a random matrices
set. These are $\boldsymbol{A_{i}}$'s in eq. (\ref{eq:Lin Mod})
and nearby expressions, and while watching at their structure a strategy
springs for calculation of those products in a binomial way, one who
breaks up them in a stochastic part and another that accounts for
memory. Stripping in a first step the deterministic matricial skeleton
out of scalar random values it is possible to classify and calculate
the many products that have been put forward. It is also possible
without lose of information formulate the products of random variables
and sum up them as well as to assign such probabilistic functions
to corresponding matrices' products, or more properly speaking to
classes of products. These algorithm-like procedure is condensed and
used in sections 5 and 6 to write needed norms of eq. (\ref{eq:Bound X_g+1})
in an operative way, eqs. (\ref{eq:Bin M_g,j 2}) and (\ref{eq:Eigenvalues M_g,j})
in the light of the general formula for above mentioned products,
summarized in eq. (\ref{eq:Binom J_m 2}).

Once is finished this quasi-computational narrative about products
of matrices through a binomial like expansion of them a discussion
in section 7 is carried out to evaluate the number of useful products
which contributes to the result. The representation chosen suggests
a fractal distribution of those surviving contributors over the whole
set of all possible products in the expansion. A criterion, among
many available in the field, is given to characterize the fractality,
(\emph{i.e.} dimension), of this \emph{binomial set} and two equivalent
ways to count elements are implemented. First an empirical one which
possesses the purpose of methodology goodness illustration, (in the
very optimistic aim of not being this problem an isolated one but
the tip of iceberg for similar random maps or dynamical models), and
second a extension of the former originated in number theory which
gives an exact answer to the question of fractal dimension and content.
Effectivity of a few steps approach is granted when compared both
results for dimension quest, since $s_{e}$ $=$ $0.691$$\pm$$0.003$
and $s_{t}$ $=$ $0.694$.

The original motivation of this study was a careful reading of a paper
from Cabrera \emph{et al. \cite{Cabr-2021} }in which they employ
linearization of DRM to explain the origin of power spectrum of some
theories of turbulence. At some point of their research they face
how $\boldsymbol{P_{g,-1}}$ behaves and $\|\boldsymbol{P_{g,-1}}\|_{2}$
grows when $g$ increases, as well in what way eigenvalues of $\boldsymbol{M_{g,j}}$
$\equiv$ $\boldsymbol{P_{g,-1}^{\dagger}}\boldsymbol{P_{g,-1}}$
reflect a subjacent fractal structure generated by all coefficients
of the former product $\boldsymbol{P_{g-1,-1}}$. We here have just
suggested a procedure which should allow them to go ahead with a full
stochastic simulation and a complete description of equation (\ref{eq:Bound X_g+1}),
for arbitrary $g$. That is, one formulation and run involving also
$\boldsymbol{P_{g,i}}$, $i=0,\dots,g$ and not only the operator
$\boldsymbol{P_{g,-1}}$ attached to initial conditions $\vec{X}_{0}$.

\subsection*{Acknowledgement}

This research did not receive any specific grant from funding agencies
in the public, commercial, or not-for-profit sectors.

\end{document}